# MONITORING COMPLIANCE WITH GOVERNMENTAL AND INSTITUTIONAL OPEN ACCESS POLICIES ACROSS SPANISH UNIVERSITIES

## Seguimiento del cumplimiento de las políticas de acceso abierto gubernamental e institucional en universidades españolas

**Reme Melero, David Melero-Fuentes, and Josep-Manuel Rodríguez-Gairín**

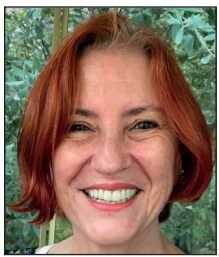

**Remedios Melero**, PhD in Chemical Sciences from the *University of Valencia*, is a researcher at the *Institute of Agrochemistry and Food Technology*, of the *Higher Council for Scientific Research* (*CSIC*) and editor of the scientific journal *Food Science and International Technology*. She is editor of the *Directory of Open Access Journals* (*DOAJ*). Member of the scientific committee of *Redalyc* and *SciELO* Spain. Coordinator of the Spanish Open Access to Science working group, participates on a national project related to research data and open science. She is a member of the *Maredata* thematic network, and a partner in the *Foster+* project *(Facilitate Open Science Training for European Research)* for the promotion of European open science policies and training in related topics.
*https://orcid.org/0000-0002-1813-8783*

*Consejo Superior de Investigaciones Científicas (CSIC), Instituto de Agroquímica y Tecnología de Alimentos (IATA)*
*Catedrático Agustín Escardino 7. 46980 Paterna (Valencia), Spain*
*rmelero@iata.csic.es*

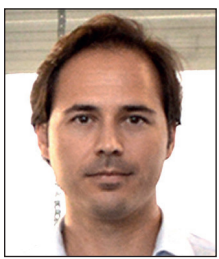

**David Melero-Fuentes**, PhD in Documentation (European mention) from the *Catholic University of Valencia San Vicente Mártir* (*UCV*), is a professor and secretary of the *Institute of Documentation and Information Technology* (*Indotei*) at the same university. During his doctorate he was a beneficiary of a predoctoral stay of the *UCV* in the *Department of Bibliometrics* of the *University of Vienna*. Its main fields of activity are evaluation of research, informetrics, information retrieval and analysis of social networks.
*https://orcid.org/0000-0002-4610-3000*

*Universidad Católica de Valencia San Vicente Mártir (UCV)*
*Instituto de Documentación y Tecnologías de la Información (Indotei)*
*Quevedo, 2. 46001 Valencia, Spain*
*david.melero@ucv.es*

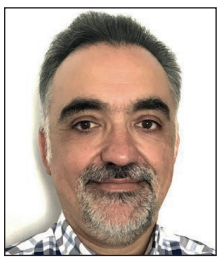

**Josep-Manuel Rodríguez-Gairín** holds a PhD in Documentation from the *University of Barcelona* (*UB*) and a lecturer in the *School of Library and Information Sciences* at the *UB*, where he also coordinates computer rooms and advises on technological aspects. He has carried out the infrastructure of projects such as *Digital journals of librarianship and information science* (*Temaria*); *BiD* journal: *University texts of librarianship and documentation*; *Information matrix for the evaluation of journals* (*MIAR*); *Online resources to carry out research works* (*Alehoop*); *Directory of experts in information handling* (*EXIT*); *International registry of authors-Links to identify scientists* (*IraLIS*), *Chronology of Spanish Documentation* (*CroDoc*), etc. He is a member of *Ciepi*, *ThinkEPI* and the technical council of the *E-LIS* repository. Founder of the *Kronosdoc* company, dedicated to the consultancy and development of documentary management programs.
*http://orcid.org/0000-0001-8375-7911*

*Universitat de Barcelona, Facultat de Biblioteconomia i Documentació*
*Melcior de Palau, 140. 08014 Barcelona, Spain*
*rodriguez.gairin@ub.edu*

**Abstract**

Universities and research centers in Spain are subject to a national open access (OA) mandate and to their own OA institutional policies, if any, but compliance with these requirements has not been fully monitored yet. We studied the degree of OA archiving of publications of 28 universities within the period 2012-2014. Of these, 12 have an institutional OA mandate, 9 do not require but request or encourage OA of scholarly outputs, and 7 do not have a formal OA statement but are well known for their support of the OA movement. The potential OA rate was calculated according to the publisher open access policies indicated in *Sherpa/Romeo* directory. The universities showed an asymmetric distribution of 1% to 63% of articles archived in repositories that matched those indexed by the *Web of Science* in the same period, of which 1% to 35% were OA and the rest were closed access. For articles on work carried out with public funding and subject to the Spanish *Science law*, the percentage was similar or slightly higher. However, the analysis of potential OA showed that the figure could have reached 80% in some cases. This means that the real proportion of articles in OA is far below what it could potentially be.

**Resumen**

Las universidades y centros de investigación en España están sujetos a un mandato de acceso abierto (OA) nacional, de acuerdo con la *Ley de la ciencia* en su artículo 37, y con sus propias políticas institucionales de libre acceso, si las hubiese, pero todavía no existe un seguimiento regular del cumplimiento de estos requisitos. En este estudio se analiza el grado de depósito de las publicaciones de 28 universidades en el período 2012-2014. De éstas, 12 tienen un mandato institucional de OA, 9 recomiendan el depósito en OA de los resultados académicos, y 7 no tienen una declaración formal de OA, pero son conocidas por su apoyo a este movimiento. La ratio de OA potencial se calculó de acuerdo con las políticas de autoarchivo de las revistas, extraídas del directorio *Sherpa/Romeo*. Las universidades mostraron una distribución asimétrica en cuanto al depósito de los artículos, variando entre el 1% al 63%, tomando como referencia los artículos indexados por la *Web of Science* en el mismo período. De éstos, entre 1% a 35% estaban en OA y el resto en acceso restringido. El porcentaje de artículos depositados que declaraban tener una financiación de fondos estatales y por tanto sujetos a la *Ley de la ciencia* española, fue similar al obtenido cuando se consideró la producción total de cada una de las instituciones, o ligeramente superior. Sin embargo, el análisis del acceso abierto potencial mostró que la cifra podría haber alcanzado el 80% en algunos casos. Esto significa que la proporción real de artículos en acceso abierto está muy por debajo de lo que potencialmente podría ser.

**Palabras clave**

Acceso abierto; Cumplimiento; Mandatos *open access*; OA; Monitorización del acceso abierto; Políticas *open access*; Acceso abierto potencial; España; Universidades; *Ley de la Ciencia*.

## 1. Introduction

"Open access literature is digital, online, free of charge, and free of most copyright and licensing restrictions" (**Suber**, 2004), so anyone can benefit from reading and using the research. There are two routes to achieving OA to scholarly publications: the green and the gold routes *BOAI* (2002). Gold OA means publication in journals that are freely accessible with or without article publishing charges (APCs). Green OA means publishing in a journal and self-archiving the published articles in an OA repository; the version that can be deposited depends on the publishers' posting policies and authors' rights.

Universities, research institutions and funders increasingly require open access to scholarly outputs. For example, the *Research Councils UK* (2014) require open access to peer-reviewed research articles resulting from projects funded by them, and allow embargoes to deposit between 6 and 12 months. The *Higher Education Funding Council for England* (2014) also requires articles and conference proceedings to be openly available and deposited in an institutional or subject repository in order to be eligible for submission to the *Research Excellence Framework* (*REF*). This requirement began to be applied to journal articles and conference proceedings accepted for publication after 1 April 2016.

The *Wellcome Trust* (2005), the *National Institute of Health* (2008), the *National Science Foundation* (2015), the *European Research Council* (2007) and the *H2020* program of the *European Commission* (2016) have OA policies whereby authors who receive funding from them are required to deposit their publications in a repository, and failure to comply may lead to withdrawal of funds.

The *University of Liège* (2007), which is a model for other institutions, bases the research evaluation exercise on the materials deposited in its repository. *Harvard University* (2008), the *Massachusetts Institute of Technology* (2009), the *University of Southampton* (**Sale**, 2006) and the *Univer-* *sidade do Minho* (2005) have also been pioneers in declaring OA policies for their publications and establishing mechanisms to monitor self-archiving.

However, despite the implementation of regulations and policies that favor OA, compliance with self-archiving in repositories is still far from 100% (**Swan** *et al.*, 2015). Even monitoring compliance is not an easy task, because it is sometimes difficult to determine the total scientific and academic output of a university, and to determine whether and where it is available in OA. As an approximation, *Scopus* and the *Web of Science* (*WoS*) are used to search papers published by an institution, though it is known and accepted that the results of the search do not include all publications. Current research information systems (CRIS) are another source of information on the scholarly outputs of a university. They contain information on the research projects that are underway and the metadata of the publications arising from them, and tend to be linked to and interoperable with the institutional repositories (**Ribeiro**; **De-Castro**; **Mennielli**, 2016). Therefore, the universities themselves are the most suitable agents for monitoring and ensuring that publications are deposited in the institutional repositories as part of an OA policy, thus avoiding dispersion of the work on several websites (**Harnard**, 2015).

The reasons for self-archiving may vary according to the area of work and the type of repository. According to the results of the *PEER* (*Publishing and the Ecology of European Research*) project (**Spezi** *et al.*, 2013), the three reasons most identified by the authors for archiving in an institutional repository were "required by employer", "invited by the repository" and "invited by a librarian". For subject repositories, the reasons were "voluntarily", "invited by publisher" or "required by research funder". The *PEER* project also found differences between disciplines. Voluntary self-archiving is greater in the physical sciences, mathematics, social sciences and humanities than in biomedical sciences. Natural and health sciences prefer the gold road, whereas physics, mathematics, social sciences and humanities prefer the green road. **Eger**, **Scheufen** and **Meierrieks** (2015) reached the similar conclusions from a survey of authors from German universities.

Open access to scholarly publications favors visibility, increases the impact of research, increases the number of readers, and breaks economic barriers between countries and communities. However, in practice authors do not exercise their right and/or duty to self-archive, because of lack of knowledge of publisher policies on self-archiving, fear of infringing copyright, not knowing how to archive, not having time and not trusting the repositories (**Frass**; **Cross**; **Gardner**, 2014).

One way to overcome these obstacles is to establish an institutional policy that encourages self-archiving, acknowledges the support received by researchers and uses the material in repositories as a source for evaluating teaching staff. Indeed, the staff assessment policy at the *University of Liège* only takes into account what is archived in their institutional repository, even if it is under embargo or closed access (*University of Liège*, 2007).

## 1.1. Previous findings regarding OA papers online and OA policy compliance

Several studies indicate the proportion of publications available in OA worldwide by countries and by disciplines. **Björk** *et al.* (2010) took a random sample of 1,837 articles published in 2008 and found that 20.4% were OA: 8.5% on publisher's websites and 11.9% on other websites. **Archambault** *et al.* (2013) reported that 43% of the articles published between 2004 and 2011 indexed in *Scopus* were OA: 33% by the green road and the hybrid road (articles published in toll access journals but with optional OA by payment of an APC) and 10% by the gold road. **Chen** (2014), in a study of articles published in 2013 and indexed in *Scopus*, found that 37.8% were available in OA on journal websites, personal web pages, institutional repositories, social networks or other websites. **Kahbsa** and **Giles** (2014) used *Google Scholar* and *Microsoft Academic Search* applied mathematical methods to analyze all the articles published between 2004 and 2011, and found that 24% were in OA. **Jubb** *et al.* (2015) published a report entitled *Monitoring the transition to open access*, commissioned to analyze the status of OA compliance in UK universities either by publishing in OA journals or by posting in institutional and subject repositories. The results revealed that 19% of articles, compared to those indexed in *Scopus*, were posted in repositories and available online, but this figure included articles that were already openly accessible on the publishers' sites immediately on publication; excluding these, the estimate would have been 9%. In view of these results, it is clear that the data obtained depend on the sources of reference, how they are obtained and the discipline involved.

> Most universities request archiving of the author's peer-reviewed final draft or the publisher's version of record, in agreement with the version specified in the Spanish *Science law*, but there are also cases in which versions are not specified. This lack of specification leads to uncertainty that does not facilitate self-archiving

In 2011 the *European Commission* announced a pilot initiative on OA to peer-reviewed research articles resulting from projects funded under the *Seventh Framework Program (FP7)* for a total of seven areas of knowledge (*European Commission*, 2011). In the *Horizon 2020* program (*H2020*), the pilot has become a mandate that covers all areas of knowledge, with embargoes of 6 and 12 months, respectively, for science, technology, engineering and mathematics (STEM) and for social sciences and humanities (*European Commission*, 2016). To measure compliance with the OA policy of *FP7* and *H2020*, *OpenAIRE* (a European project created to support the implementation of the OA policies of the *European Commission* and the *European Research Council*) harvests the metadata of the papers deposited in institutional repositories whose research is funded by the *European Commission*. To this end, the *OpenAIRE* guidelines use a project identifier field for standardized specification of the funding agency, in this case the *European Commission* (*OpenAIRE*, 2015). In fact, trials of this type have been carried out with the *Fundação para a Ciência e a Tecnologia* (*FCT*) and the *Wellcome Trust*. As of 6 October 2016, the statistics provided by *OpenAIRE* indicated that 64% of publications of projects under the *FP7* open access pilot were in OA (*OpenAIRE*, 2016a). According to data collected by *OpenAIRE*, the *Fundação para a Ciência e a Tecnologia* (*FCT*) reached a figure of 92.6%, very similar to the 90.8% reached by the *European Grid Initiative* community (*OpenAIRE*, 2016b). Spanish institutional repositories have implemented these guidelines for European projects, but there is still no standard format for expressing information on national projects. The Spanish authorities are expected to publish a recommendation on project identification in the very near future, and it will then be possible to monitor national projects in a similar way to European projects.

In May 2016 the *Schweizerischer Nationalfonds* published a report on the monitoring of its mandate policy on the green and gold roads (**Gutknecht** *et al.*, 2016). The report covered the period 2013-2015 and the sources of reference were initially the *WoS* and *Scopus*. The first analysis compared the publications in these citation databases that contained in the acknowledgments information on funding by the *Swiss National Science Foundation* (*SNSF*) with all OA publications, including those deposited in repositories, in OA journals and on personal websites. The level of coincidence was only 20%. After a validation using the digital object identifier (DOI) and searching in other sources (*DOAJ*, *PubMed*, *PubMed Central*, *OpenAIRE* and the *Astrophysics Data System*), the figure reached 56%, of which 27% corresponded to publications deposited in OA repositories; this value is close to that obtained by **Borrego** (2015) for OA publications in Spain.

### 1.2. The OA context at national level in Spain

The new Spanish *Law 14/2011, on science, technology and innovation* (hereafter the Spanish *Science law*), which contains an article on OA (article 37), was passed in 2011 (*España*, 2011; *Fecyt*, 2014). In accordance with this law, publications arising from projects funded by the general state budget should be deposited in an OA repository as soon as possible and no later than twelve months after the official date of publication. In fact, this requirement was already included in the latest calls of the Spanish *National science, innovation and technology plan* (*Mineco*, 2013; 2017). This requirement is in line with other policies such as those of H2020 (*European Commission*, 2016). However, according to Point 6 of Article 37 of the *Law*, the authors may be exempted from depositing if they have reached rights assignment agreements with third parties (in most cases with publishers). In addition, universities and research centers have established their own institutional policies for making the scholarly outputs of their staff available in OA, and for preserving them. Therefore, self-archiving of the scholarly publications of Spanish universities should be favored by both the Spanish *Law* and by any institutional mandates or recommendations that are in place.

**Borrego** (2015) conducted a study to estimate compliance with the Spanish *Science law* within the period 2011-2014 using the articles published in 2012 with government funding obtained. Taking a random sample of all the projects, he found that of all the articles of 2012 indexed in the *WoS*, at least 58.4% were available on the internet in OA journals, repositories or other websites. Of these articles, 23.8% were published in OA journals and 21.8% were archived in repositories; of the latter, most were in the subject repositories *Arxiv* or *PubMed*. In total, an average of 14.4% of the articles resulting from publicly funded research were available in institutional repositories, and the distribution between institutions was asymmetric. This figure is close to the 12.4% obtained by the *Pasteur4OA Project* (*Open access policy alignment strategies for European Union research*) for institutions with an institutional mandate (**Swan** *et al.*, 2015).

Bearing in mind these precedents and the lack of information on monitoring of and compliance with OA policies in Spain, the aim of the present study was to analyze the degree of compliance with OA policies by Spanish universities from two points of view:

(a) Institutional compliance (for universities with institutional OA policies that require or encourage self-archiving of the scholarly publications of their staff), and (b) compliance with Article 37 of the Spanish *Science law*, which requires published papers resulting from projects funded by the Spanish government to be deposited as soon as possible and no later than 12 months after their publication. This analysis covered the period 2012-2014, in which the mandate of the *Law* was in force and at the time of the study the embargo period of 12 months laid down by the *Law* had expired. Moreover, taking into account both approaches to analyze if there are any synergy effects between the policies.

### 2. Samples and methods

To monitor compliance with OA institutional policies and compliance with the Spanish *Law*, we have to identify the total number of peer-reviewed articles subject to the policy, the total number of full-text OA articles in the repository, and the number of embargoed full-text items that will become OA at a later date, as suggested by **Picarra** (2015).

Compliance with institutional or governmental OA policies was analyzed in the period 2012-2014 to ensure that any embargoes had expired, on the assumption that in most journals the embargo period was less than 24 months (the Spanish *Science law* establishes a period of 12 months after publication). We used the *Bielefeld Academic Search Engine* (*BASE*) and its application programming interface (API) (*Bielefeld University Library*, 2016a; 2016b, respectively) to obtain the XML files with the records of the articles deposited in institutional repositories in this period. To obtain the published works subjected to the policy, we used the databases included in the *WoS Core Collection* (*Clarivate Analytics*, 2017).

### 2.1. The Spanish universities studied

We drew up a list of the 25 Spanish universities listed in the *Melibea* directory (*Acceso Abierto*, 2016) that have OA policies (mandate or recommendation). We also added 7

Table 1. Spanish universities included in this study and their open access policies, if any

| University | Acronym | Open access policy | Berlin declaration signatory |
|---|---|---|---|
| Univ. Carlos III de Madrid | UC3M | Mandate/Requirement | Yes |
| Univ. Complutense de Madrid | UCM | Mandate/Requirement Mandate/Requirement | Yes |
| Univ. de Burgos | UBU | Mandate/Requirement | Yes |
| Univ. de Huelva | UHU | Mandate/Requirement | Yes |
| Univ. de Las Palmas de Gran Canaria | ULPGC | Mandate/Requirement | Yes |
| Univ. de León[1] | UNILEON | Mandate/Requirement | Yes |
| Univ. Nacional de Educación a Distancia | UNED | Mandate/Requirement | Yes |
| Univ. Politécnica de Madrid | UPM | Mandate/Requirement | Yes |
| Univ. Rey Juan Carlos | URJC | Mandate/Requirement | Yes |
| Univ. Rovira i Virgili[1] | URV | Mandate/Requirement | No |
| Univ. Autònoma de Barcelona | UAB | Mandate/Requirement | Yes |
| Univ. de Barcelona | UB | Mandate/Requirement | Yes |
| Univ. Oberta de Catalunya | UOC | Mandate/Requirement | Yes |
| Univ. Politècnica de Catalunya | UPC | Mandate/Requirement | Yes |
| Univ. CEU Cardenal Herrera | CEU | Recommend/Encourage | Yes |
| Univ. de Alcalá | UAH | Recommend/Encourage | Yes |
| Univ. de Cantabria | UNICAN | Recommend/Encourage | Yes |
| Univ. de Extremadura[1] | UEX | Recommend/Encourage | No |
| Univ. de Málaga[1] | UMA | Recommend/Encourage | Yes |
| Univ. Politécnica de Cartagena | UPCT | Recommend/Encourage | Yes |
| Univ. Politécnica de Valencia | UPV | Recommend/Encourage | Yes |
| Univ. de Girona | UdG | Recommend/Encourage | Yes |
| Univ. de Lleida | UdL | Recommend/Encourage | Yes |
| Univ. de Vic | UVIC | Recommend/Encourage | Yes |
| Univ. Pompeu Fabra | UPF | Recommend/Encourage | Yes |
| Univ. Autónoma de Madrid | UAM | - | Yes |
| Univ. de Alicante | UA | - | Yes |
| Univ. del País Vasco | EHU | - | No |
| Univ. Jaume I | UJI | - | Yes |
| Univ. Pablo de Olavide | UPO | - | Yes |
| Univ. Pública de Navarra | UPNA | - | Yes |
| Univ. de València | UV | - | Yes |

1. Excluded from this study (see Methods)

other universities that do not have an explicit policy but advocate in favor of OA and/or have signed the *Berlin declaration on open access to knowledge in the sciences and humanities* (*Open access at the Max Planck Society*, 2003). From these, we excluded the *University of Extremadura* and the *University of Málaga* because their institutional repositories contained only articles published in the universities' own journals. Also the *University of León* because it indexed the deposit date instead of the publication date in the field *dc:date* or *dc:year*; and the *Universitat Rovira i Virgili* because its repository only contained doctoral theses at the time when the data were gathered. We analyzed a total of 28 institutions: twelve with an OA mandate, nine that encourage OA and seven without an institutional OA policy, all of which were subject to Article 37 of the Spanish *Science law* (Table 1).

Table 2 shows some features of the OA policies and the corresponding type according to the university OA policy classification proposed by **Shieber** and **Suber** (2015):

Type 1. The policy grants the institution certain non-exclusive rights to future research articles published by faculty staff. This type of policy typically offers a waiver option or opt-out for authors. It also requires depositing in the repository.

Type 2. The policy requires faculty to retain certain non-exclusive rights when they publish future research articles. Whether or not it offers a waiver option for authors, it requires depositing in the repository.

Table 2. Some features of the institutional OA policies of studied universities

| Institution | Effective from YYYY-MM-DD | OA policy | Allow faculty to opt-out of the requirement? | Versions of papers | When to deposit | Allowed embargo | Copyright reservation | Type of policy |
|---|---|---|---|---|---|---|---|---|
| UNED | 2014-07-14 | Green OA mandate and recommended gold OA | No opt-out of deposit but opt-out of immediate OA, case by case | Author's peer-reviewed final draft. Publisher's version of record | At the time of publication | 12 months | The policy grants the institution certain non-exclusive rights to future research articles published by faculty | 1 |
| UC3M | 2010-01-08 | Green OA mandate | Both opt-outs of deposit and of immediate OA | Author's peer-reviewed final draft. Publisher's version of record | Unspecified | Estipulated by the publisher | Unspecified | 4 |
| CEU | 2014-03-06 | Recommended green and gold OA | Not applicable | Unspecified | At the time of publication | Estipulated by the publisher | No copyright reservation | 5 |
| UCM | 2014-05-27 | Green OA mandate | No opt-out of deposit but opt-out of immediate OA | Unspecified | As soon as possible | Estipulated by the publisher | University recommends to avoid exclusively copyright transfers to allow selfarchiving | 3 |
| UAH | 2013-03-21 | Recommended Green OA | Unspecified | Unspecified | Unspecified | Estipulated by the publisher | Unspecified | 5 |
| UBU | 2014-03-31 | Green OA mandate | Both opt-outs of deposit and of immediate OA | Unspecified | As soon as possible | Estipulated by the publisher | Unspecified | 4 |
| UNICAN | 2012-07-24 | Recommended green and gold OA | Not applicable | Author's peer-reviewed final draft. Publisher's version of record | At the time of acceptance or publication | Estipulated by the publisher | Unspecified | 5 |
| UHU | 2015-02-27 | Green OA mandate and recommended gold OA | No opt-out of deposit but opt-out of immediate OA | Version allowed by the publisher | At the time of acceptance | Between 6 and 12 months | University recommends to avoid exclusively copyright transfers to allow selfarchiving | 3 |
| ULPGC | 2015-10-08 | Green OA mandate | No opt-out of deposit but opt-out of immediate OA | Unspecified | As soon as possible | 12 months | No copyright reservation | 3 |
| UPCT | 2011-04-13 | Recommended green and gold OA | Not applicable | Author's peer-reviewed final draft. Publisher's version of record | Unspecified | Unspecified | Unspecified | 5 |
| UPM | 2010-10-28 | Green OA mandate and recommended gold OA | No opt-out of deposit but opt-out of immediate OA | Unspecified | Unspecified | Estipulated by the publisher | Unspecified | 3 |
| UAB | 2012-04-25 | Green OA mandate and recommended gold OA | No opt-out of deposit but opt-out of immediate OA | Author's peer-reviewed final draft. Publisher's version of record. Unrefereed preprint | At the time of publication | 6 months | No copyright reservation | 3 |
| UB | 2012-01-01 | Green OA mandate and recommended gold OA | No opt-out of deposit but opt-out of immediate OA | Unspecified | At the time of publication | 6 months | No copyright reservation | 3 |
| UdG | 2012-01-09 | Recommended green and gold OA | Not applicable | Author's peer-reviewed final draft. Publisher's version of record | At the time of publication | 6 months | Unspecified | 5 |
| UdL | 2012-05-30 | Recommended green and gold OA | Not applicable | Author's peer-reviewed final draft. Publisher's version of record | At the time of publication | 12 months | Unspecified | 5 |
| UVIC | 2012-10-16 | Recommended green and gold OA | Not applicable | Unspecified | At the time of aceptance | 6 months | Unspecified | 5 |
| UOC | 2010-10-06 | Green OA mandate and recommended gold OA | No opt-out of deposit but opt-out of immediate OA | Author's peer-reviewed final draft. Publisher's version of record | At the time of acceptance | 12 months | The policy grants the institution certain non-exclusive rights to future research articles published by faculty | 1 |
| UPC | 2009-10-07 | Green OA mandate and recommended gold OA | No opt-out of deposit but opt-out of immediate OA | Author's peer-reviewed final draft. Publisher's version of record | As soon as possible | 6 months for those funded by national projects | Unspecified | 3 |
| UPV | 2011-07-21 | Recommended green and gold OA | Not applicable | Author's peer-reviewed final draft, Publisher's version of record | At the time of publication | Estipulated by the publisher | No copyright reservation | 5 |
| UPF | 2011-04-06 | Recommended green and gold OA | Not applicable | Publisher's version of record, pre-print | At the time of publication | 6 months | No copyright reservation | 5 |

1. According to the Stuart Shieber and Peter Suber classification (**Shieber**; **Suber**, 2015).

Type 3. The policy seeks no rights at all, but requires depositing in the repository. If the institution already has permission to make a work OA, then it makes it OA from the moment of deposit. Otherwise, the deposit will be "dark" (non–OA) until the institution can obtain permission to make it OA. During the period of dark deposit, at least the metadata will be OA.

Type 4. The policy seeks no rights at all and does not require dark deposits. It requires repository depositing and OA, but only when the author's publisher permits them.

Type 5. The policy does not require OA in any sense, but merely requests or encourages it.

Type 6. The policy does not require OA in any sense, but asks faculty to "opt in" to a policy under which they are expected to deposit their work in the repository and authorize it to be OA.

‘ The ratio of articles listed in the *WoS* that were published and deposited in institutional repositories (Deposit-INST) in the period 2012-2014 ranged from 1% to 62% ’

## 2.2. Obtaining the articles deposited in institutional repositories

The metadata of the articles published in scholarly journals during the period 2012-2014 and deposited in repositories of the universities studied were harvested from *BASE* (*Bielefeld University Library*, 2016a). The XML files were obtained using the API of *BASE*. We executed an equation for each university with the following parameters:

(a) search function (func=PerformSearch);

(b) repository queried, named according to the *BASE* nomenclature (target=questioned<internal_name>);

(c) fields consulted (query=<queryterm>&(...));

(d) maximum number of bibliographic records returned by the equation (hits=<number>); and

(e) fields returned by the query for each record (fields=<field1, field2... >).

The required fields in the search were:

*dc:title* (title of the document)

*dc:creator* (examples of a creator include a person, an organization, or a service)

*dc:contributor* (an entity responsible for making contributions to the content of the resource)

*dc:date* (year of publication)

*dc:identifier* (example formal identification systems include the URI, URL and DOI)

*dc:relation* (the DC element relation can be used to indicate different kinds of relations between several metadata records)

*dc:rights* (open, embargoed or closed access to the document), and

*dc:type* (type of document).

If necessary, these fields might be repeated for each element (for example, the *dc:identifier* field would appear three times if it contains an URI, an URL and a DOI). The repositories collected met the *OpenAIRE* guidelines (*OpenAIRE*, 2015), so in the description of the metadata they used the syntax of the guidelines. In the absence of a national standard for specifying projects financed by the Spanish government, some repositories provided this information in the *dc:relation* field using their own criteria. In some cases by entering the project code, in others by entering the text of the acknowledgments of the articles, and in others using a syntax similar to that of European projects.

info:*eu-repo/grantAgreement/Funder/FundingProgram/ProjectID*

As we were only interested in articles, we introduced in the search equation the restriction of articles in the document type, which according to the codes of the *BASE* API corresponds to 121 (*doctype*:121).

For example, to retrieve the articles deposited in the repository of the *University of Alicante* (*UA*) that were published in the period 2012-2014, we designed the following equation:

https://api.base-search.net/cgi-bin/BaseHttpSearchInterface.fcgi?func=PerformSearch&target=ftunivalicante&query=dcyear:[2012+TO+2014]&doctype:121&fields=dctitle,dccreator,dccontributor,dcdate,dcidentifier,dcrelation,dcrights,dctype

In this case, target refers to the related institution name (ftunivalicante), *dc:year* to the period of publication (from 2012 to 2014), *doc:type* to the type of publication, and 121 code to articles.

The records retrieved for each university in XML format were exported and tabulated in a *MS Excel* spreadsheet. The records of each university were then manually checked.

### 2.3. Obtaining the articles indexed in the *WoS*

For each of the 28 universities, we implemented a search equation in the *WoS* to retrieve the articles published in the period 2012-2014. Each search equation was composed of two search lines. One line retrieved all the articles published by the university and the other retrieved only the articles financed by government funding bodies.

Both search sets queried (a) the Organization Data field (labeled OG) to select each university and (b) the Address field (AD), where the variant names of each university were consulted. In addition, the search set that retrieved only the articles financed by public funding queried the fields Funding Agency (FO), Grant Number (FG) and Funding Text (FT) with the following terms referring to government funding: Spain, Spanish, MINECO, MEC, MINCINN, Ministerio, Espana, CSIC, ISCIII, "Carlos III Health Institute", CICYT, "Consejo Superior de Investigaciones Cientificas", "Consolider Program", FICYT, FIS, "Fondo de Investigacion Sanitaria", "Fondo de Investigaciones Sanitarias", INIA, "Iniciativa Ingenio", "Instituto Carlos III", "Instituto de Salud Carlos III", MICINN, "Ministry of Economy and Competitiveness", "Ministry of Education", "Ministry of Education and Science", "Ministry of Science and Innovation" and "Ministry of Science and Technology".

The records retrieved for each university were then downloaded and tabulated in *MS Excel*.

> The ratio of *WoS* articles deposited in repositories ranged from 2% to 76% for articles that acknowledged funding by Spanish government bodies (Deposit-GOV)

### 2.4. Calculating the rate of OA compliance by individual institutions

We took as a reference of the published papers of an institution the articles indexed in the *WoS* databases in the period 2012-2014. For each university, we consulted each year and in the three years analyzed, we determined the total number of articles and, of these, the number of government-funded articles. We considered government-funded articles as a fraction of the whole output published by a university indexed in *WoS*. We identified the number of articles (funded and not funded) that were indexed in both *WoS* and

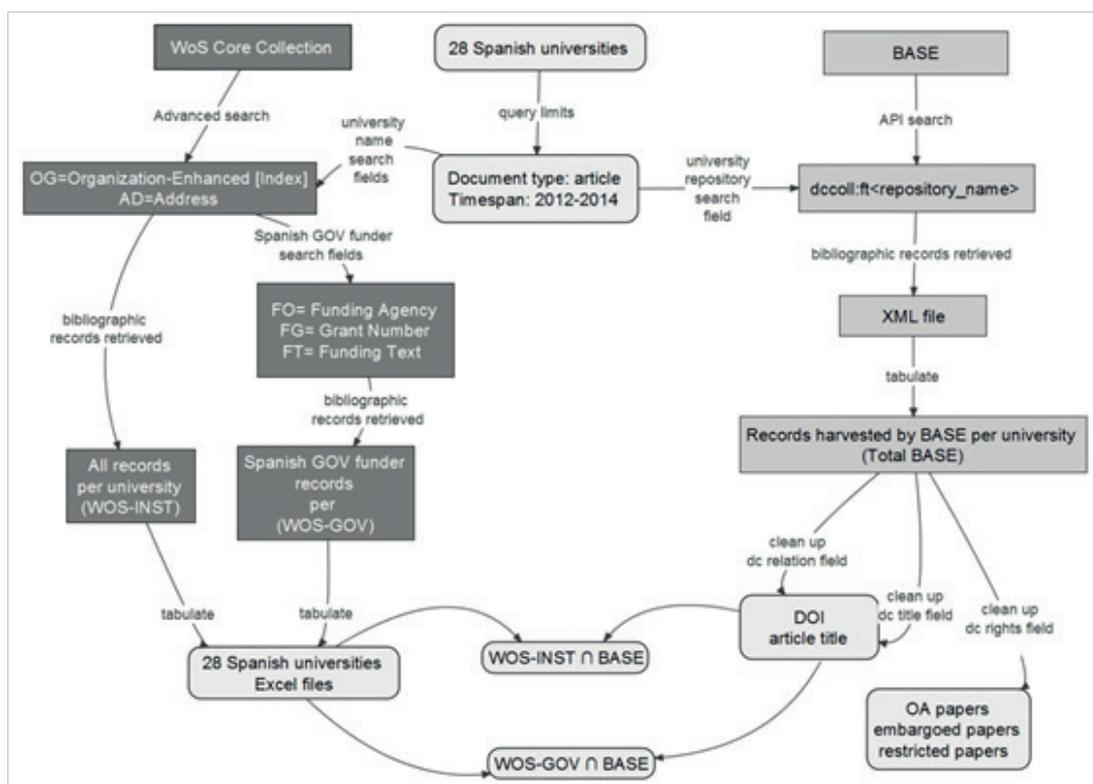

Figure 1. Flow chart of the search for articles indexed in the *WoS* and the data harvested from *BASE* to obtain the ratio of articles deposited in institutional repositories to the total published in 2012-2014 by 28 Spanish universities.

*BASE* through an algorithm that consulted the DOIs and the titles of the articles (Figure 1). If a DOI or title matched, the record was considered a positive match if the year of publication also coincided (note that the search already coincides with the type of document and one of the signatory institutions). The mismatched titles were sorted alphabetically and revised manually. Full stops were eliminated from the titles and the double (or longer) spaces were normalized to one.

Duplicates within an institution were eliminated, either because they were different versions of the same article or because there had been some duplication in the deposit by members of different departments. We did not take into account possible duplicates in repositories of different universities as a result of collaboration between researchers, because we were measuring compliance by institutions rather than overall compliance by country.

> The universities with the highest ratio of depositing showed the highest ratio of closed access (except *UPM*) but also the highest ratios of access to the rest of the articles

The information in the *dc:rights* field was used to distinguish between articles that were OA, subject to embargo or closed according to the *OpenAIRE* guidelines. Previously, we checked that this field was correctly populated, but the time in which the repositories began to use it differed, so it appears in blank in some records.

The Institutional compliance index (ICI) was calculated using the following equation:

ICI= (Total compliant articles archived in the period 2012-2014 (OA+embargoed))/(Total number of articles subject to the institutional policy in the period (2012-2014)indexed in WoS) ×100

Where the numerator is the number of papers harvested from *BASE* and indexed in *WoS*, and the denominator is the total number of papers retrieved and indexed in *WoS* corresponding to a university within the period 2012-2014.

The Governmental compliance index (GCI) was calculated using the equation

GCI= (Total GOV-compliant articles archived in the period 2012-2014 (OA+embargoed))/(Total number of articles subject to the GOV-policy in the period (2012-2014)indexed in WoS) ×100.

where the numerator is the number of papers harvested from *BASE* that acknowledge funding by a government entity and are indexed in the *WoS*, and the denominator is the total number of papers corresponding to the university within the period 2012-2014 that were retrieved and indexed in the *WoS* and acknowledged funding by a government body.

We also included embargoed papers in the equations because data were retrieved in 2016, two years after their publication, which was sufficient time to go beyond the permitted embargo of 12 months stated in the Spanish *Science law*.

### 2.5. Calculating the potential self-archiving index

The Potential self-archiving index (PAI) was defined as the proportion of articles subject to an institutional or governmental policy that can be deposited according to the archiving policies of the journals in which they are published and the color assigned by *Sherpa/Romeo* (a directory of publisher copyright policies and self-archiving): green, blue, yellow and white (*Sherpa/Romeo*, 2016a). We used the API of *Sherpa/Romeo* (*Sherpa/Romeo*, 2016b) to calculate the potential proportion of articles in green journals (which allow self-archiving in pre- and post-print version), in blue journals (which allow self-archiving in the post-print version), in yellow journals (which allow self-archiving in the pre-print version) and in white journals (which do not allow self-archiving). For each university, the calculation took into account all the articles indexed in the *WoS* that were subject to an institutional and governmental policy:

PAI= (Articles indexed in WoS in the period (2012-2014) classified by Romeo colours)/(Total number of articles subject to the policy in the period (2012-2014) indexed in WoS) ×100

### 3. Results and discussion

Of the 28 universities studied, 12 have an OA mandate, 9 request or recommend OA, and 7 have no formal OA institutional statement but are well known for their support to the OA movement. Table 2 shows some features of the OA institutional policies and the corresponding type according to the OA policy classification proposed by **Shieber** and **Suber** (2015). Policies of types 1 and 3 are the strongest, because they require archiving without exemption; if authors do not have permission, the deposit remains "dark" (non–OA) until the institution obtains permission to make it open. Type 1 corresponds to the model of *Harvard University* (**Suber**, 2015), adopted by the Spanish *National University of Distance Education* (*UNED*) and approved by its senate in July 2014. Type 3 requires archiving but not necessarily in open access, deposits might be embargoed or closed, but at least the metadata of the articles are openly available. The authors of this classification do not recommend policies of type 4 because they allow recalcitrant publishers to opt-out at will, and type 5 is a mere recommendation.

In summary, of the 28 universities studied, 12 have an institutional OA mandate, but only two do not allow to opt-out of archiving. Of the rest, nine do not require but request or encourage OA to scholarly outputs, and seven do not have a formal OA statement but are well known for their support to the OA movement.

Most universities request archiving of the author's peer-reviewed final draft or the publisher's version of record, in agreement with the version specified in the Spanish *Science* *law*, but there are also cases in which versions are not specified. This lack of specification leads to uncertainty that does not facilitate self-archiving. In addition, if it is not specified when the articles must be archived, it could be delayed indefinitely: the more loopholes there are, the weaker the policies are.

The date of application of the policies ranges from 2009 to 2015, so the ones approved toward the end of the study period (2012-2014) will have had practically no effect on archiving. This is the case of the *UNED* (2014), the *CEU Cardenal Herrera University* (2014), the *UCM* (2014), the *University of Burgos* (*UBU*, 2014), the *University of Huelva* (*UHU*, 2015) and the *University of Las Palmas de Gran Canaria* (*ULPGC*, 2015).

### 3.1. Papers published by universities in the period 2012-2014

According to the data of the *WoS*, scholarly publications in the period 2012-2014 showed on average an annual growth of 5.5% and no changes were observed in the order of the universities by volume of articles published over the period (Figure 2A). In the *WoS*, the proportion of all articles that were funded by the Spanish government in the period 2012-2014 was above 50% for all the universities studied (Figure 2B), except for the *UOC* (25%) and the *UNED* (39%), which are distance learning universities. Rather than comparing universities by the total volume of articles published, which depends on their size, we show the average number of articles produced per tenured research staff member per year in the period 2012-2014 (Figure 2A), according to the data provided by the observatory *Actividad Investigadora en la Universidad Española* (*Iuene*, 2016). The *Pompeu Fabra University* (*UPF*), a small- to medium-sized university (approx. 8,000 students) had the highest ratio, with 3.63 articles per staff member, followed by the large universities *UAB* and the *UB* (> 40,000 students), with two articles per staff member.

### 3.2. Ratio of articles deposited in institutional repositories

The ratio of articles listed in the *WoS* that were published and deposited in institutional repositories (Deposit-INST) in the

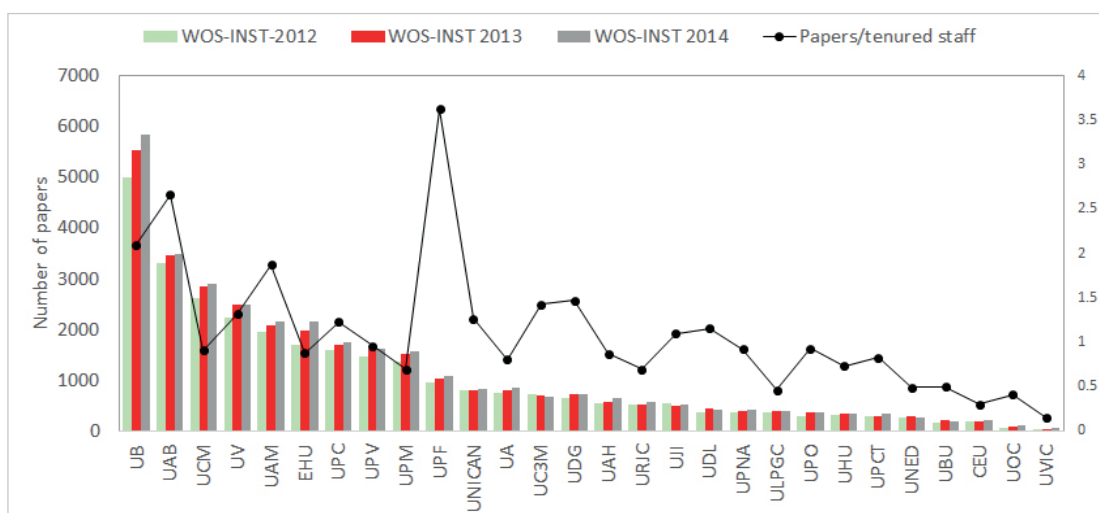

2A

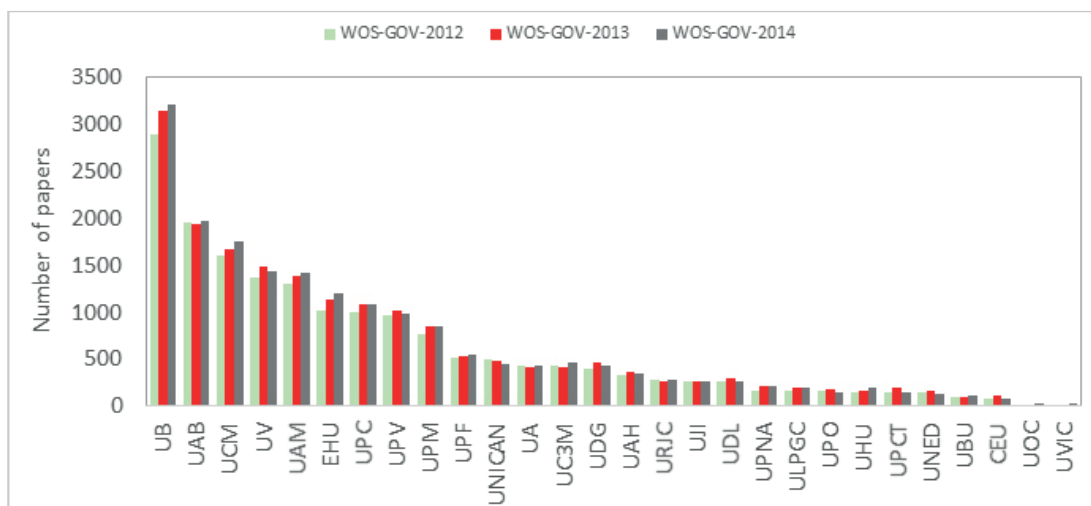

2B

Figure 2. (A) Number of articles published by the universities studied between 2012 and 2014. (B) Articles published by the universities studied between 2012 and 2014 that acknowledged funding by government bodies (Source: *WoSCC*, 2016). Continuous line shows the average number of articles per tenure staff in the same period (Source: *Iuene*, 2016).

period 2012-2014 ranged from 1% to 62% (Figure 3A). The universities with the highest ratios were small- to medium-sized universities (the *University of Vic* (*UVIC*) and the *Jaume I University* (*UJI*), with approx. 5000 and 15,000 students, respectively. In the case of the *UVIC*, only 136 articles were published and indexed in the *WoS* in the three years (see Table 4), so self-archiving or mediated deposit were facilitated by the low workload. At the *Jaume I University* (*UJI*), deposits are mainly mediated by the library, so the effect of voluntary deposits is minimized. In terms of the archiving ratio, the *Universidad de Alicante*, the *Universitat Politècnica de Catalunya* (*UPC*), the *Universidad Politécnica de Valencia* (*UPV*) and the *Universidad Politécnica de Madrid* (*UPM*) follow these universities. The *UPC* and the *UPM* have institutional OA mandates that state that authors cannot opt-out of depositing but can opt-out of immediate OA (see Table 2).

The ratio of *WoS* articles deposited in repositories ranged from 2% to 76% (Figure 3B) for articles that acknowledged funding by Spanish government bodies (Deposit-GOV).

Figure 3C shows the total number of papers indexed in the *WoS* in the period 2012-2014 (Total *WoS*) for each of the universities studied, and those that acknowledge funding by a governmental body (Total *WoS*-GOV). In almost all universities studied, the number of papers harvested from the repositories was very low compared with the number of published articles indexed in the *WoS*, with the exception of the *UPC*, the *UA*, the *UJI*, the *University of Lleida* (*UdL*), the *UNED*, the *UHU* and the *UVIC*. Of these, the *UPV*, the *UA*, the *UJI* and the *UVIC* had the highest ratios. The *UNED* had a very low ratio, but the number of papers deposited in its repository was higher than the number indexed in the *WoS*, so many of them were from other sources.

According to the data from the *WoS Core Collection*, the percentage of published articles deposited in repositories is slightly higher for government-funded articles (on average 55% of the total) than for the rest.

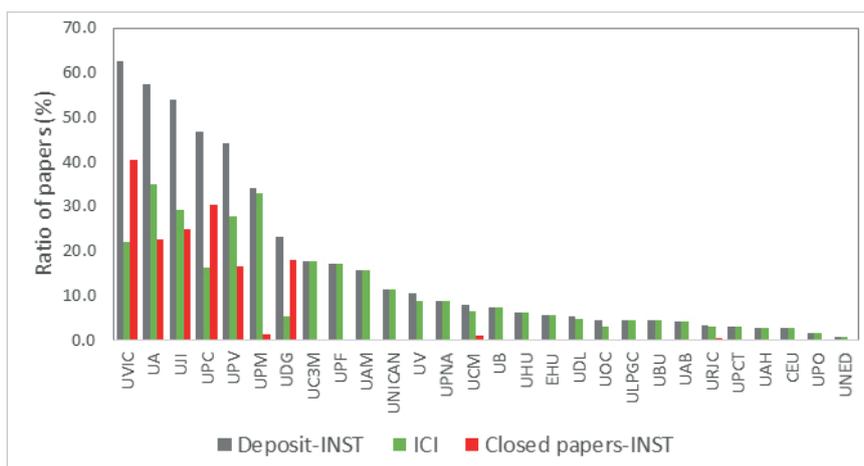

3A

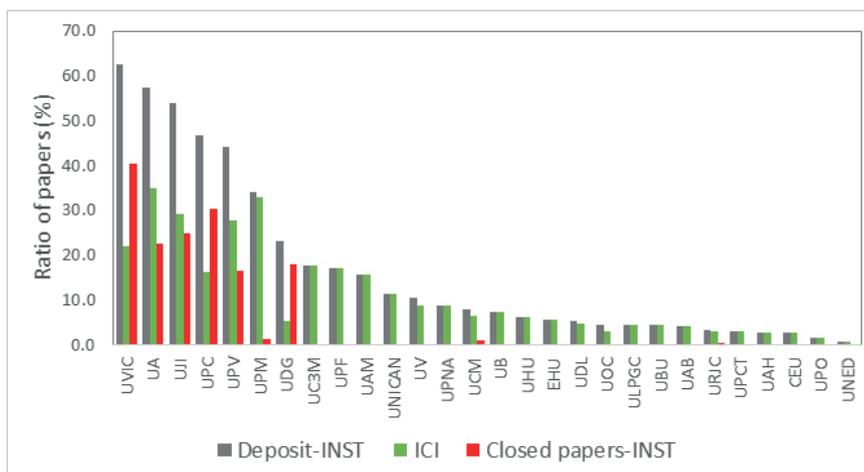

3B

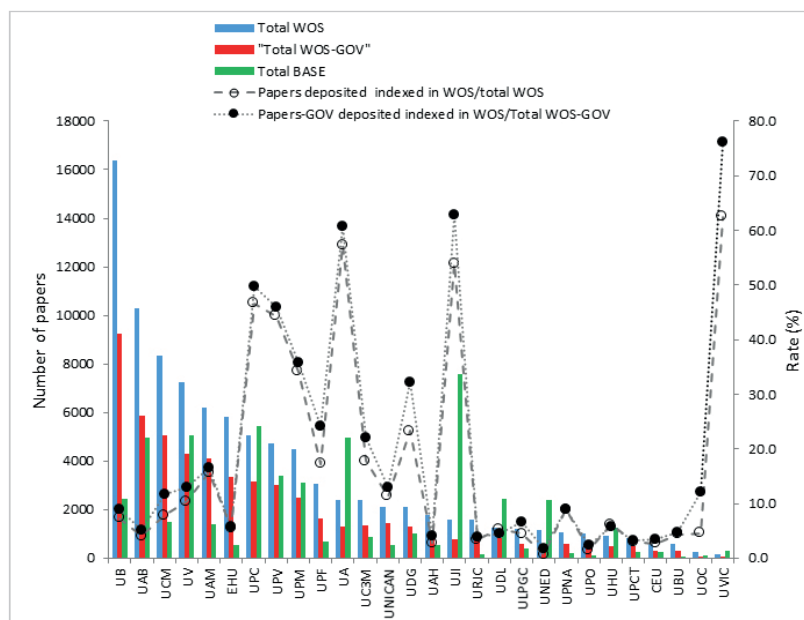

3C

Figure 3. (A) Ratio of all articles indexed in the WoS that were deposited in the institutional repository (Deposit-INST) in relation to ICI and closed papers (Closed papers-INST). (B) Ratio of papers acknowledging funding by government bodies (Deposit-GOV) that were deposited (Deposit-GOV) in relation to GCI and closed papers (Closed papers-GOV). (C) Number of papers indexed in WoS (Total *WoS*), and of those total papers funded by government entities (Total *WoS*-GOV) compared to the ones archived in repositories (Total *BASE*) and percentage of matching with articles indexed in WoS.

We classified the universities into three groups regarding the type of institutional policy: Group 1 represents universities with no institutional policy, Group 2 universities that encourage depositing and Group 3 universities that have a mandate. All three groups are subject to the Spanish *Science law*.

If we represent the depositing ratios and the *ICI* and *GCI* compliance indices separately for each group, we obtain the adjustments of Figures 4A and 4B. In view of these results, there seem to be no differences between the three groups, so it is not possible to distinguish the effect of the institutional policies from the requirement established in Article 37 of the Spanish *Science law*. The data on OA availability will be dealt with in the following section.

### 3.3. Ratio of OA articles archived in institutional repositories

Not all deposited articles in institutional repositories were OA. Table 3 shows the results obtained from *BASE* for each university, identified by the metadata *dc:rights* as openAccess, embargoedAcccess and closedAccess, terms from the *info:eu-repo-Access-Terms* vocabulary (*Surfnet*, 2013). According to this vocabulary, "embargoed access" means the resource is closed access until released for OA on a certain date, and "closed access" means that the item is not available in the public internet and is also known as "toll-gated access".

The *dc:rights*-empty column shows the number of records that could not be assigned to any of these categories. The *UOC* and the *Universitat de València* (*UV*) showed high figures of 32% and 47%, respectively, which could have a considerable effect on the calculation of the articles that are available in OA in their repositories.

For each university we calculated the percentage of articles indexed in the *WoS* that were deposited in repositories and the percentage that were OA or closed access (Table 3). If we assume that the ones with an embargo will potentially be OA after a certain time, the OA ratio for each university would be the sum of OA and embargoed articles.

In 23 of the universities studied, most of the articles archived in their repositories are available in OA (Table 3), with the exception of the following institutions, which had a significant percentage of closed papers: the *UA* (16%), the *UJI* (17%), the *UPV* (29%), the *UPC* (40%), the *UdG* (47%) and the *UV* (47% of blank *dc:rights* metadata). Of these, the *UPC* has an OA mandate type 3 policy (Table 2) so the authors cannot opt-out of deposit but opt-out of immediate OA. The percentage of articles deposited, compared to the ones indexed in *WoS*, ranged from 0.7% to 62.5%. The universities with the highest ratio of depositing (*UA*, *UdG*, *UJI*, *UPC*, *UPV* and *UVIC*) were also those with the highest percentage of closed access papers. Nevertheless, these universities still had the highest proportion of articles available in OA.

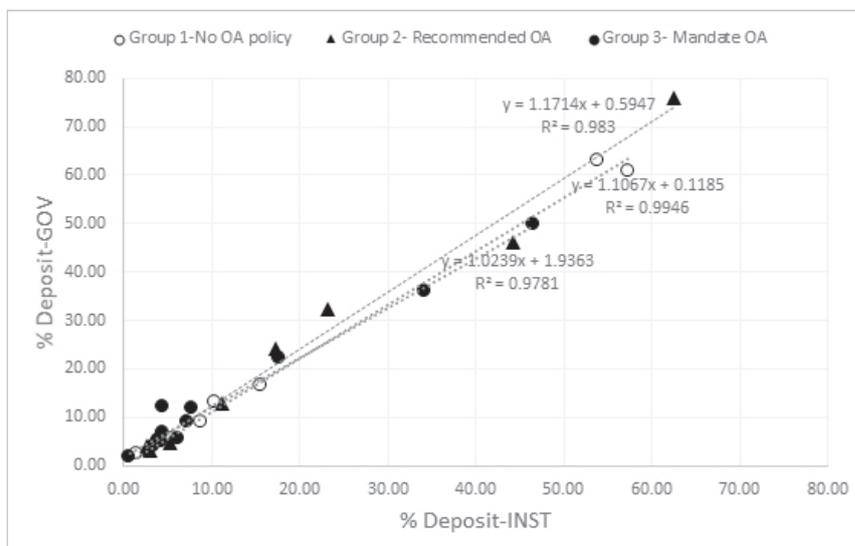

4A

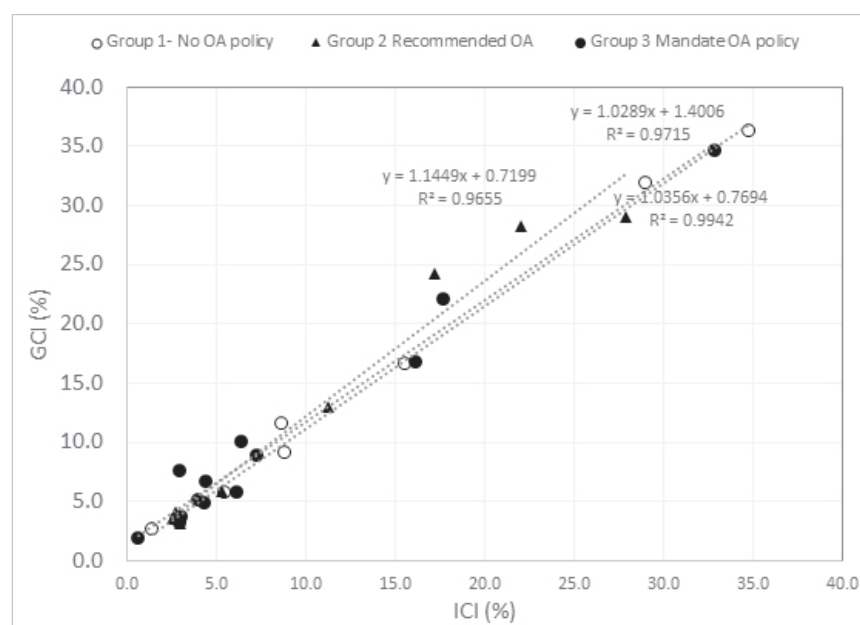

4B

Figure 4. (A) Adjustment of the percentage of articles deposited in institutional repositories (Deposit-INST) in relation to those acknowledging funding by government bodies (Deposit-GOV). (B) Adjustment of Institutional compliance index (ICI) in relation to the Government compliance index (GCI). Group 1: Universities that do not have an open access policy but are subject to the Spanish *Science law*. Group 2: Universities that have an open access recommendation policy and are subject to the Spanish *Science law*. Group 3: Universities that have an open access mandate policy and are subject to the Spanish *Science law*.

Table 3. Number of articles harvested by *BASE* during the period 2012-2014, classified as open, embargoed and closed papers according to the element *dc:rights*

| University | Number of harvested articles | | | | |
| --- | --- | --- | --- | --- | --- |
| | Total | Open access | Embargoed | Closed access | *Dc:rights*-empty* |
| CEU | 238 | 238 | 0 | 0 | 0 |
| UA | 4,978 | 4,159 | 11 | 808 | 0 |
| UAB | 4,979 | 4,907 | 13 | 59 | 0 |
| UAH | 550 | 549 | 0 | 1 | 0 |
| UAM | 1,392 | 1,381 | 11 | 0 | 0 |
| UB | 2,454 | 2,422 | 30 | 2 | 0 |
| UBU | 59 | 59 | 0 | 0 | 0 |
| UC3M | 853 | 819 | 33 | 1 | 0 |
| UCM | 1,474 | 1,126 | 160 | 142 | 46 (3%) |
| UdG | 995 | 507 | 19 | 469 | 0 |
| UdL | 2,428 | 2,324 | 15 | 6 | 83 (3.4%) |
| UHU | 1,265 | 1,265 | 0 | 0 | 0 |
| UJI | 7,588 | 6,287 | 2 | 1,298 | 1 (0.01%) |
| ULPGC | 399 | 399 | 0 | 0 | 0 |
| UNED | 2,377 | 2,377 | 0 | 0 | 0 |
| UNICAN | 525 | 525 | 0 | 0 | 0 |
| UOC | 119 | 81 | 0 | 0 | 38 (32%) |
| UPC | 5,448 | 3,295 | 0 | 2,153 | 0 |
| UPCT | 264 | 264 | 0 | 0 | 0 |
| UPF | 704 | 701 | 3 | 0 | 0 |
| UPM | 3,096 | 2,829 | 74 | 179 | 14 (0.5%) |
| UPNA | 198 | 188 | 10 | 0 | 0 |
| UPO | 113 | 91 | 0 | 0 | 22 (0.4%) |
| UPV | 3,372 | 2,178 | 224 | 970 | 0 |
| EHU | 562 | 561 | 1 | 0 | 0 |
| URJC | 141 | 127 | 0 | 14 | 0 |
| UV | 5,072 | 2,678 | 0 | 0 | 2394 (47%) |
| UVIC | 322 | 184 | 9 | 129 | 0 |

*Dc:rights* empty means the element does not contain any information.

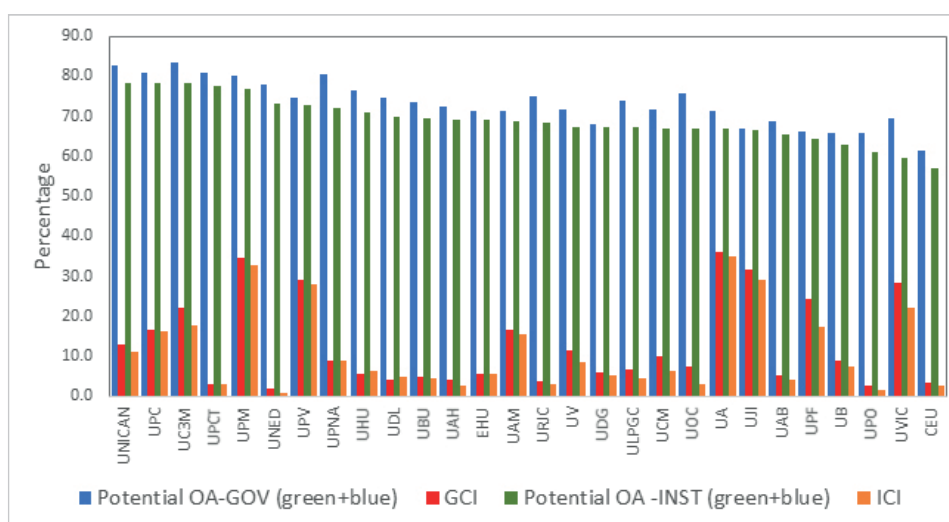

Figure 5. Comparison of real (measured as ICI and GCI) and potential open access according to whether the articles had been published in green or blue journals based on the *Sherpa/Romeo* taxonomy. The graph includes all articles according to affiliation (INST) and those acknowledging funding by government bodies (GOV)

## 3.4. Calculating depositing and OA of articles financed through national projects

Table 4 shows the number of articles indexed in the *WoS* that acknowledge funding from government bodies and the number of these that are deposited. The behavior is similar to that of the previous section. The universities with the highest ratio of depositing (*UA*, *UdG*, *UJI*, *UPC*, *UPM*, *UPV* and *UVIC*) showed the highest ratio of closed access (except the *UPM* with 1.3%), but also the highest ratios of access to the rest of the articles. In the rest of the universities, closed access was zero or very low. The range of values (2% to 36%) was slightly higher than that calculated taking into account the articles of the whole institution, regardless of funding.

The *UPC* and the *UPM* were included in a study reported by the *Pasteur4OA* project to analyze OA policies in Europe in the period 2011-2013 (**Swan** *et al.*, 2015). In comparison with our findings, the rate of OA papers is slightly higher: 32.8% and 16.2% for the *UPM* and the *UPC*, respectively, compared with 26.2% and 13.9% for the *UPM* and the *UPC*, respectively. This difference could be due to the period used to calculate that percentage, which in our case was 2012-2014, one year ahead.

‘ The results show that the potential OA of published papers ranged between 60% and 80%, well above the average ’

## 3.5. Potential OA by universities according to journal archiving policies

Each of the articles gathered from the *WoS* was given a *Sherpa/Romeo* color according to the policy of the journal in which it had been published. The colors (green, blue, yellow and white) indicate whether the journal allows self-archiving and at what stage of the publication process. From all the records of the *WoS* for the period 2012-2014 for each university, we obtained the data shown in Table 6.

Green and blue indicate that the publisher's version or the

Table 4. Total articles published and indexed in *WoS* by the corresponding university from 2012 to 2014, number of total deposits and closed papers, and ICI (Institutional compliance index)

| University | WoS total | Total deposited | | Closed | | OA | Embargoed | ICI (%) |
|---|---|---|---|---|---|---|---|---|
| | | n | % | n | % | | | |
| CEU | 575 | 15 | 2.6 | 0 | | 15 | 0 | 2.6 |
| UA | 2,416 | 1,384 | 57.3 | 542 | 22.4 | 836 | 6 | 34.9 |
| UAB | 10,276 | 416 | 4.0 | 1 | 0.0 | 412 | 3 | 4.0 |
| UAH | 1,759 | 48 | 2.7 | 0 | | 48 | 0 | 2.7 |
| UAM | 6,197 | 970 | 15.7 | 0 | 0.0 | 960 | 10 | 15.7 |
| UB | 16,363 | 1,198 | 7.3 | 1 | 0.0 | 1185 | 12 | 7.3 |
| UBU | 571 | 25 | 4.4 | 0 | | 25 | 0 | 4.4 |
| UC3M | 2,393 | 425 | 17.8 | 0 | | 400 | 25 | 17.8 |
| UCM | 8,363 | 654 | 7.8 | 82 | 1.0 | 442 | 97 | 6.4 |
| UdG | 2,110 | 490 | 23.2 | 377 | 17.9 | 100 | 13 | 5.4 |
| UdL | 1,238 | 66 | 5.3 | 3 | 0.2 | 53 | 7 | 4.8 |
| UHU | 925 | 57 | 6.2 | 0 | | 57 | 0 | 6.2 |
| UJI | 1,596 | 861 | 53.9 | 396 | 24.8 | 464 | 0 | 29.1 |
| ULPGC | 1,191 | 53 | 4.5 | 0 | | 53 | 0 | 4.5 |
| UNED | 1,140 | 8 | 0.7 | 0 | | 8 | 0 | 0.7 |
| UNICAN | 2,111 | 238 | 11.3 | 0 | | 238 | 0 | 11.3 |
| UOC | 264 | 12 | 4.5 | 0 | | 8 | 0 | 3.0 |
| UPC | 5,040 | 2,349 | 46.6 | 1532 | 30.4 | 817 | 0 | 16.2 |
| UPCT | 835 | 25 | 3.0 | 0 | | 25 | 0 | 3.0 |
| UPF | 3,076 | 529 | 17.2 | 0 | | 527 | 2 | 17.2 |
| UPM | 4,464 | 1,527 | 34.2 | 58 | 1.3 | 1407 | 59 | 32.8 |
| UPNA | 1,059 | 94 | 8.9 | 0 | | 84 | 10 | 8.9 |
| UPO | 1,022 | 15 | 1.5 | 0 | | 15 | 0 | 1.5 |
| UPV | 4,716 | 2,088 | 44.3 | 774 | 16.4 | 1121 | 193 | 27.9 |
| EHU | 5,828 | 323 | 5.5 | 0 | | 323 | 0 | 5.5 |
| URJC | 1,566 | 54 | 3.4 | 6 | 0.4 | 48 | 0 | 3.1 |
| UV | 7,227 | 752 | 10.4 | 0 | | 630 | 0 | 8.7 |
| UVIC | 136 | 85 | 62.5 | 55 | 40.4 | 24 | 6 | 22.1 |

Reme Melero, David Melero-Fuentes, and Josep-Manuel Rodríguez-Gairín

Table 5. Total articles published and indexed in *WoS* with mention of funding by a government body (*WoS*-GOV) during the period 2012-2014, number of total deposits and closed papers and GCI (Government compliance index).

| University | Total papers *WoS*-GOV | Total deposits | | Closed papers | | OA papers | Embargoed papers | GCI (%) |
|---|---|---|---|---|---|---|---|---|
| | | n | % | n | % | | | |
| CEU | 283 | 10 | 3.5 | 0 | 0.0 | 10 | 0 | 3.5 |
| UA | 1,297 | 788 | 60.8 | 318 | 24.5 | 467 | 3 | 36.2 |
| UAB | 5,879 | 298 | 5.1 | 1 | 0.0 | 295 | 2 | 5.1 |
| UAH | 1,049 | 42 | 4.0 | 0 | | 42 | 0 | 4.0 |
| UAM | 4,104 | 678 | 16.5 | 0 | | 672 | 6 | 16.5 |
| UB | 9,246 | 821 | 8.9 | 1 | 0.0 | 809 | 11 | 8.9 |
| UBU | 314 | 15 | 4.8 | 0 | | 15 | 0 | 4.8 |
| UC3M | 1,330 | 293 | 22.0 | 0 | | 282 | 11 | 22.0 |
| UCM | 5,036 | 587 | 11.7 | 61 | 1.2 | 411 | 90 | 9.9 |
| UdG | 1,297 | 419 | 32.3 | 343 | 26.4 | 67 | 9 | 5.9 |
| UdL | 824 | 38 | 4.6 | 3 | 0.4 | 30 | 3 | 4.0 |
| UHU | 510 | 29 | 5.7 | 0 | | 29 | 0 | 5.7 |
| UJI | 799 | 503 | 63.0 | 248 | 31.0 | 254 | 0 | 31.8 |
| ULPGC | 568 | 38 | 6.7 | 0 | | 38 | 0 | 6.7 |
| UNED | 454 | 8 | 1.8 | 0 | | 8 | 0 | 1.8 |
| UNICAN | 1,446 | 188 | 13.0 | 0 | | 188 | 0 | 13.0 |
| UOC | 66 | 8 | 12.1 | 0 | | 5 | 0 | 7.6 |
| UPC | 3,170 | 1,575 | 49.7 | 1,046 | 33.0 | 529 | 0 | 16.7 |
| UPCT | 501 | 16 | 3.2 | 0 | 0.0 | 16 | 0 | 3.2 |
| UPF | 1,616 | 391 | 24.2 | 0 | 0.0 | 389 | 2 | 24.2 |
| UPM | 2,473 | 887 | 35.9 | 28 | 1.1 | 830 | 27 | 34.7 |
| UPNA | 597 | 54 | 9.0 | 0 | | 51 | 3 | 9.0 |
| UPO | 486 | 12 | 2.5 | 0 | | 12 | 0 | 2.5 |
| UPV | 2,989 | 1,376 | 46.0 | 509 | 17.0 | 743 | 124 | 29.0 |
| EHU | 3,357 | 191 | 5.7 | 0 | | 191 | 0 | 5.7 |
| URJC | 834 | 32 | 3.8 | 1 | 0.1 | 31 | 0 | 3.7 |
| UV | 4,298 | 558 | 13.0 | 0 | | 494 | 0 | 11.5 |
| UVIC | 46 | 35 | 76.1 | 22 | 47.8 | 11 | 2 | 28.3 |

accepted reviewed post-print can be deposited. Because this is the condition of Article 37 of the Spanish *Science law*, we took the sum of the two as potential OA, i.e. what the OA ratio would have been if the authors had made full use of the possibility of self-archiving. The results show that the potential OA of published papers ranged between 60% and 80%, well above the average.

### 3.6. Potential OA of articles with government funding according to journal archiving policies

As in the previous section, this calculation was made with the articles that acknowledged funding from one of the bodies mentioned in the methodology. The results are shown in Table 7.

Both for all the articles (INST) and for those acknowledging funding from government bodies (GOV), the real percentages were below the potential ones indicated by the *Sherpa/Romeo* color codes (Figure 6).

Comparing the real data with the potential one for articles published in green and blue journals, we found differences of 30% to 70%, showing that depositing is still far lower than it could be. If repositories took advantage of this, the archiving rate could rise enormously, but in order to reach those figures, repository managers and/or librarians should track the articles published by their staff, and authors should take care to keep at least the version accepted for publication.

### 3.7. Limitations of the study

The limitations of this study are related to the sources used, which may have influenced the results obtained:

- The *WoS* limits the sources of reference to journals indexed in this database, so a future study should use other databases, such as *Scopus* (which includes *Medline*). However, the *WoS* includes the most important scientific publications in each subject area (**Ruiz-Pérez**; **Delgado-López-Cózar**; **Jiménez-Contreras**, 2006; **Ruiz-Pérez**; **Delgado-López-Cózar**, 2013), while *Scopus* over-represents peripheral literature in the scientific communication system (**López-Illescas**; **De-Moya-Anegón**; **Moed**, 2008; **Bartol** *et al.*, 2014).

Tabla 6. Potential self-archiving index (PAI) of total papers indexed in *WoS* corresponding to Spanish universities (with or without funding statement) classified by colors according to the journal in which they were published and the *Sherpa/Romeo* journal taxonomy (green, blue, yellow and white, see definitions in section 2.5 in methods).

| University | Published papers in 2012-2014 | | | | White | | PAI (%) | | |
|---|---|---|---|---|---|---|---|---|---|
| | Total | Green | Blue | Yellow | n | % | Green | Blue | Yellow |
| CEU | 575 | 284 | 44 | 118 | 88 | 15.3 | 49.4 | 7.7 | 20.5 |
| UA | 2416 | 1443 | 176 | 341 | 265 | 11.0 | 59.7 | 7.3 | 14.1 |
| UAB | 10,276 | 6,193 | 529 | 1,971 | 997 | 9.7 | 60.3 | 5.1 | 19.2 |
| UAH | 1,759 | 1,129 | 89 | 256 | 174 | 9.9 | 64.2 | 5.1 | 14.6 |
| UAM | 6,197 | 3,947 | 325 | 880 | 627 | 10.1 | 63.7 | 5.2 | 14.2 |
| UB | 16,363 | 9,400 | 915 | 3326 | 1,828 | 11.2 | 57.4 | 5.6 | 20.3 |
| UBU | 571 | 361 | 36 | 59 | 75 | 13.1 | 63.2 | 6.3 | 10.3 |
| UC3M | 2,393 | 1,770 | 100 | 322 | 50 | 2.1 | 74.0 | 4.2 | 13.5 |
| UCM | 8,363 | 5,030 | 581 | 1,218 | 888 | 10.6 | 60.1 | 6.9 | 14.6 |
| UdG | 2,110 | 1,296 | 127 | 327 | 269 | 12.7 | 61.4 | 6.0 | 15.5 |
| UdL | 1,238 | 804 | 62 | 189 | 83 | 6.7 | 64.9 | 5.0 | 15.3 |
| UHU | 925 | 580 | 78 | 96 | 67 | 7.2 | 62.7 | 8.4 | 10.4 |
| UJI | 1,596 | 975 | 87 | 232 | 204 | 12.8 | 61.1 | 5.5 | 14.5 |
| ULPGC | 1,191 | 735 | 68 | 199 | 83 | 7.0 | 61.7 | 5.7 | 16.7 |
| UNED | 1,140 | 705 | 127 | 133 | 65 | 5.7 | 61.8 | 11.1 | 11.7 |
| UNICAN | 2,111 | 1,564 | 92 | 216 | 120 | 5.7 | 74.1 | 4.4 | 10.2 |
| UOC | 264 | 134 | 43 | 58 | 5 | 1.9 | 50.8 | 16.3 | 22.0 |
| UPC | 5,040 | 3,767 | 183 | 602 | 254 | 5.0 | 74.7 | 3.6 | 11.9 |
| UPCT | 835 | 621 | 28 | 101 | 39 | 4.7 | 74.4 | 3.4 | 12.1 |
| UPF | 3,076 | 1,734 | 247 | 749 | 206 | 6.7 | 56.4 | 8.0 | 24.3 |
| UPM | 4,464 | 3,150 | 272 | 452 | 240 | 5.4 | 70.6 | 6.1 | 10.1 |
| UPNA | 1,059 | 707 | 58 | 182 | 49 | 4.6 | 66.8 | 5.5 | 17.2 |
| UPO | 1,022 | 537 | 87 | 188 | 100 | 9.8 | 52.5 | 8.5 | 18.4 |
| UPV | 4,716 | 3,226 | 212 | 563 | 411 | 8.7 | 68.4 | 4.5 | 11.9 |
| EHU | 5,828 | 3,792 | 232 | 730 | 723 | 12.4 | 65.1 | 4.0 | 12.5 |
| URJC | 1,566 | 983 | 88 | 298 | 95 | 6.1 | 62.8 | 5.6 | 19.0 |
| UV | 7,227 | 4,471 | 405 | 1,030 | 810 | 11.2 | 61.9 | 5.6 | 14.3 |
| UVIC | 136 | 62 | 19 | 27 | 12 | 8.8 | 45.6 | 14.0 | 19.9 |

- The effective date of the OA institutional policies is not the same for the institutions studied, so the effects of those policies are not completely comparable.
- We only considered institutional repositories to monitor OA compliance, but the Spanish *Science law* also permits self-archiving in subject repositories.
- The quality of the metadata on the source of funding was not optimal because not all records contain the information needed to describe the funding source, or it is not normalized.
- The *Sherpa/Romeo* database aims to be regularly updated, but editorial policies change even faster, so data accuracy is not 100% guaranteed.

## 4. Conclusions

Compliance with governmental and institutional OA policies varies greatly from one university to another, with an average of 11%, a maximum of 33% and a minimum of 0.7%. Compliance with Article 37 of the Spanish *Science law* is slightly higher, at 13%, 2% and 36%, respectively.

Comparing the depositing rate with the OA rate, universities with the highest percentage of OA also had the highest percentage of closed access, sometimes as much as 50%. This effect is due to the depositing of the publisher's version, which may prevent OA, since authors do not have permission to self-archive except when they publish in OA journals. According to the potential OA results, it seems that the post-print version is not being used widely for depositing, or there is a preference for the publisher's version (version of record).

Universities with a low or null closed papers ratio have a low rate of coincidence with the papers indexed in the *WoS*, and the number of papers deposited is well below the number published. However, most publications available in their repositories are OA; this might be due to an internal institutional policy of depositing only what can be openly available, as the *Carlos III University of Madrid* (*UC3M*) does, for example. In this case, repositories do not provide the option to include metadata of publications that could be deposited but remain closed until they can be released.

Table 7. Potential self-archiving index (PAI) of articles subjected to the governmental policy (PAI-GOV) that can be deposited according to the archiving policies of their journals and the color assigned by *Sherpa/Romeo* (green, blue, yellow and white, see definitions in section 2.5 in methods).

| University | Published papers in 2012-2014 funded by government projects | | | | | | PAI-GOV (%) | | |
|---|---|---|---|---|---|---|---|---|---|
| | Total | Green | Blue | Yellow | White | | Green | Blue | Yellow |
| | | | | | n | % | | | |
| CEU | 283 | 161 | 13 | 57 | 49 | 17.3 | 56.9 | 4.6 | 20.1 |
| UA | 1,297 | 865 | 61 | 158 | 152 | 11.7 | 66.7 | 4.7 | 12.2 |
| UAB | 5,879 | 3,822 | 227 | 1,001 | 651 | 11.1 | 65.0 | 3.9 | 17.0 |
| UAH | 1,049 | 735 | 24 | 141 | 124 | 11.8 | 70.1 | 2.3 | 13.4 |
| UAM | 4,104 | 2,781 | 149 | 511 | 474 | 11.5 | 67.8 | 3.6 | 12.5 |
| UB | 9,246 | 5,636 | 452 | 1,754 | 1,167 | 12.6 | 61.0 | 4.9 | 19.0 |
| UBU | 314 | 227 | 4 | 23 | 50 | 15.9 | 72.3 | 1.3 | 7.3 |
| UC3M | 1,330 | 1,090 | 22 | 128 | 29 | 2.2 | 82.0 | 1.7 | 9.6 |
| UCM | 5,036 | 3,446 | 166 | 667 | 600 | 11.9 | 68.4 | 3.3 | 13.2 |
| UdG | 1,297 | 815 | 68 | 180 | 193 | 14.9 | 62.8 | 5.2 | 13.9 |
| UdL | 824 | 592 | 24 | 116 | 72 | 8.7 | 71.8 | 2.9 | 14.1 |
| UHU | 510 | 381 | 9 | 49 | 50 | 9.8 | 74.7 | 1.8 | 9.6 |
| UJI | 799 | 530 | 6 | 107 | 138 | 17.3 | 66.3 | 0.8 | 13.4 |
| ULPGC | 568 | 398 | 22 | 70 | 49 | 8.6 | 70.1 | 3.9 | 12.3 |
| UNED | 454 | 347 | 6 | 53 | 38 | 8.4 | 76.4 | 1.3 | 11.7 |
| UNICAN | 1,446 | 1,153 | 45 | 127 | 81 | 5.6 | 79.7 | 3.1 | 8.8 |
| UOC | 66 | 48 | 2 | 13 | 1 | 1.5 | 72.7 | 3.0 | 19.7 |
| UPC | 3,170 | 2,495 | 73 | 352 | 165 | 5.2 | 78.7 | 2.3 | 11.1 |
| UPCT | 501 | 396 | 9 | 47 | 30 | 6.0 | 79.0 | 1.8 | 9.4 |
| UPF | 1,616 | 945 | 126 | 413 | 111 | 6.9 | 58.5 | 7.8 | 25.6 |
| UPM | 2,473 | 1,902 | 81 | 254 | 134 | 5.4 | 76.9 | 3.3 | 10.3 |
| UPNA | 597 | 454 | 27 | 72 | 27 | 4.5 | 76.0 | 4.5 | 12.1 |
| UPO | 486 | 296 | 23 | 87 | 59 | 12.1 | 60.9 | 4.7 | 17.9 |
| UPV | 2,989 | 2,148 | 82 | 315 | 320 | 10.7 | 71.9 | 2.7 | 10.5 |
| EHU | 3,357 | 2,325 | 66 | 365 | 501 | 14.9 | 69.3 | 2.0 | 10.9 |
| URJC | 834 | 606 | 20 | 144 | 45 | 5.4 | 72.7 | 2.4 | 17.3 |
| UV | 4,298 | 2,950 | 132 | 515 | 555 | 12.9 | 68.6 | 3.1 | 12.0 |
| UVIC | 46 | 29 | 3 | 10 | 3 | 6.5 | 63.0 | 6.5 | 21.7 |

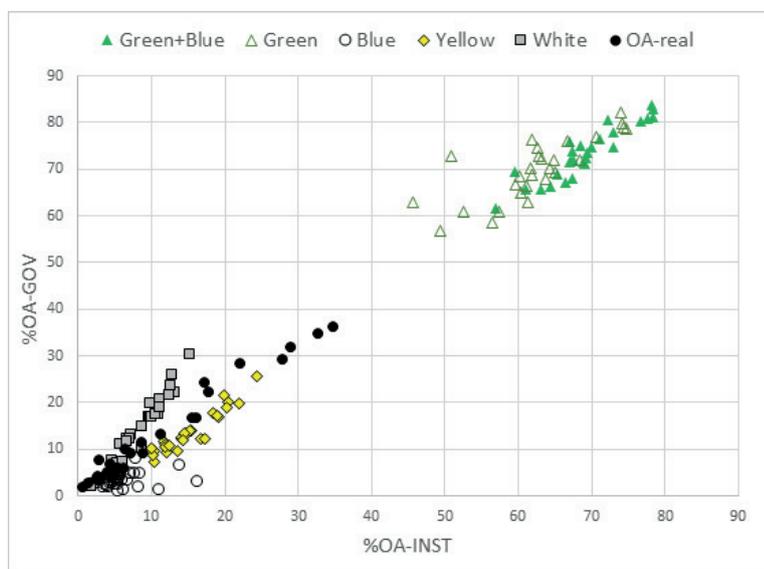

Figure 6. Percentage of real and potential open access (for INST and GOV) according to the color of the journal in which the articles were published based on the *Sherpa/Romeo* taxonomy (green, blue, yellow, white).

In view of the results, it is clear that a policy is not sufficient to encourage archiving of scholarly outputs in institutional repositories, at least in the case studied. We therefore recommend:

- To monitor compliance with institutional policies and with the mandate of the Spanish *Science law*. This would allow scheduling actions to improve archiving and indirectly them to know the behaviour of their staff with regard to sharing their publications
- The funders that require OA to the publications arising from research projects should also establish mechanisms for monitoring compliance
- To take into account the works deposited in repositories for the assessment and promotion of Faculty staff, following the model of the *University of Liège*. This seems to be a good way to increase the participation of researchers
- To use current research information systems (CRISs) as source for the scholarly outputs metadata since they provide an accurate information about their publications. With this information libraries can establish a mechanism for asking the authors for the articles directly (this system was set up at *Oregon State University*, where the archiving of works indexed in the *WoS* between 2012 and 2014 rose from 12% to 45%)
- Facilitate and encourage self-archiving will help authors to meet their obligations to their employers and funders
- All authors of articles should keep a copy of the submitted version (pre-print) and the version accepted (post-print), because they may be able to use one of these if they are not allow to use the version of record.

Finally, we have not found any clear response why the compliance is lower than expected when institutions have their own policy plus the law mandate. Lack of awareness, inertia to the status quo, threat to infringe the copyright law, and lack of incentives or recognition of open access practices, might be part of the reasons. Authors are aware that during last 2 years the efforts of librarians and project managers have favoured self-archiving, however we should compare in future works if this perception corresponds with the real situation.

> We recommend to monitor compliance with institutional policies and with the mandate of the Spanish *Science law*

## 5. Acknowledgments

The authors thank the Spanish *Ministerio de Economía y Competitividad* for funding the project CSO2014-52830-P, and the staff of the *University of Bielefeld* for allowing us to use the API of *BASE*.

## Appendix 1

| University | Set search in ADDRESS Field (AD) |
|---|---|
| CEU | "CEU Card?nal Herrera" OR "Card?nal Herrera CEU" OR "Univ* Card?nal Herrera" OR "Card?nal Herrera Univ*" OR UCH-CEU OR CEU-UCH OR UCHCEU OR CEUUCH OR "San Pablo CEU Univ*" OR "Univ* San Pablo CEU" OR "CEU San Pablo Univ*" OR "Univ* CEU San Pablo" OR CEU-USP OR USP-CEU OR CEUUSP OR USPCEU |
| UA | "Univ* Al?cant*" OR "Univ* de Al?cant" OR "Al?cant* Univ*" OR "Univ* of Al?cant*" OR "Univ* *Al?cant*" |
| UAB | "Univ* Auto* Barcelona" OR "Auto* Univ* Barcelona" OR UAB |
| UAH | "Univ* Alcala" OR UAH |
| UAM | "Univ* Auto* Madrid" OR "Univ* Auto* de Madrid" OR "Auto* Univ* Madrid" OR "Auto* Univ* of Madrid" |
| UB | "Univ* Barcelona" OR UB |
| UBU | "Univ* Burgos" OR UBU |
| UC3M | "Univ* Carlos III" OR UC3M |
| UCM | "Univ* Compluten*" OR "Compluten* Univ* Madrid" OR UCM |
| UdG | "Univ* Gerona" OR "Univ* Girona" OR *UdG* |
| UdL | "Univ* Lleida" OR "Univ* Lerida" OR UdL |
| UHU | "Univ* Huelva" OR UHU |
| UJI | "Univ* Jaume" OR "Jaume Univ*" OR "Jaume I Univ*" OR UJI (NOT Kyoto) |
| ULPGC | "Univ* Palmas Gran Canaria" OR "Univ* Las Palmas de Gran Canaria" OR "Univ* Las Palmas Gran Canaria" OR "Univ* Palmas de Gran Canaria" OR "Palmas Gran Canaria Univ*" OR "Las Palmas de Gran Canaria Univ*" OR "Las Palmas Gran Canaria Univ*" OR "Palmas de Gran Canaria Univ*" OR ULPGC |
| UNED | "Univ* Nac* Educ* Distan*" OR "Nat* Distan* Educ* Univ*" OR UNED |
| UNICAN | "Univ* Cantabria" OR UNICAN |
| UOC | "Univ* Oberta Cat*" OR "Univ* Abierta Cat*" OR "Open Univ* Cat*" OR (UOC NEAR/1 Spain) |
| UPC | "Univ* Politec* Cat*" OR "Polytech* Univ* Cat*" OR "Tech* Univ* Cat*" OR UPC |
| UPCT | "Univ* Politec* Cartagena" OR "Polytech* Univ* Cartagena" OR "Tech* Univ* Cartagena" OR UPCT |
| UPF | "Univ* Pompeu Fabra" OR "Pompeu Fabra Univ*" OR UPF |
| UPM | "Univ* Politec* Madrid" OR "Polytech* Univ* Madrid" OR "Tech* Univ* Madrid" OR UPM (NOT Malaysia) |
| UPNA | "Univ* Publ* de Navarra" OR "Univ* Publ* Navarra" OR "Nafarroako Unib* Publ*" OR "Publ* Univ* of Navarra" OR "Publ* Univ* Navarra" |
| UPO | "Univ* Pabl* de Olavide" OR "Univ* Pabl* of Olavide" OR "Pabl* de Olvaide Univ*" OR "Pabl* Olvaide Univ*" |
| UPV | "Univ* Politec* Valencia" OR "Tech* Univ* Valencia" OR "Polytech* Univ* Valencia" |
| UPV/EHU | "Univ* of Basq* Count*" OR "Univ* Basq* Count*" OR "Basq* Count* Univ*" OR "Univ* Pais Vasc*" OR "Univ* del Pais Vasc*" OR "Pais Vasc* Univ*" OR "Eus* Herri* Uniber*" OR "UPV/EHU" |
| URJC | "Univ* Rey Juan Carlos" OR "King Juan Carlos Univ*" OR URJC |
| UV | "Univ* of Valencia" OR "Univ* Valencia" OR "Valencia Univ*" OR "Univ* de Valencia" |
| UVIC-UCC | "Univ* Vic" OR "Univ* Central Cat*" OR "Central Univ* Cat*" OR UVIC OR UVIC-UCC OR UCC-UVIC |

Reme Melero, David Melero-Fuentes, and Josep-Manuel Rodríguez-Gairín

## Glossary

| Acronym | Spelled out |
|---|---|
| *BASE* | *Bielefeld Academic Search Engine* |
| *Sherpa/Romeo* | Database that shows the copyright and open access self-archiving policies of academic journals |
| *WoS* | *Web of Science* database |
| CRIS | Current research information system |
| *PEER* project | *Publishing and the Ecology of European Research* project |
| *7FP* | *Seventh Framework Program of European Union* |
| STEM | Science, technology, engineering and mathematics |
| H2020 | *Horizon 2020* program of the *European Commission* |
| FCT | *Fundação para a Ciência e a Tecnologia*, Portugal |
| *SNSF* | Schweizerischer Nationalfonds or in English Swiss National Science Foundation |
| DOI | Digital object identifier |
| *DOAJ* | *Directory of open access journals* |
| *Pasteur4OA* project | *Open access policy alignment strategies for European Union research* project |
| API | Application programming interface |
| XML | Extensible markup language |
| *WoSCC* | *Web of Science Core Collection* databases |
| DC | Dublin core schema of metadata |
| URI | Uniform resource identifier |
| URL | Uniform resource locator |
| OG | Label of organization field in *WoSCC* |
| AD | Label of address field in *WoSCC* |
| FO | Label of funding agency field in *WoSCC* |
| FG | Label of grant number field in *WoSCC* |
| FT | Label of funding text field in *WoSCC* |
| *Mineco* | Spanish *Ministry of Economy and Competitiveness* |
| *MEC* | Spanish *Ministry of Economy and Competitiveness* |
| *Mincinn* | Spanish *Ministry of Science and Innovation* |
| *CSIC* | Spanish *High Scientific Research Council* |
| *Isciii* | Spanish *Carlos III Health Institute* |
| *Cicyt* | Spanish *Comisión Interministerial de Ciencia y Tecnología* |
| *Fecyt* | Spanish *Fundación Española para la Ciencia y la Tecnología* |
| *FIS* | Spanish *Fondo de Investigacion Sanitaria* |
| *INIA* | Spanish *Instituto Nacional de Investigación y Tecnología Agraria y Alimentaria* |
| *Micinn* | Spanish *Ministry of Science and Innovation* |
| ICI | Institutional compliance index |
| GCI | Governmental compliance index |
| PAI | Potential self-archiving index |
| *IUNE* | *Obervatorio de la Actividad Investigadora en la Universidad Española* (*http://www.iune.es*) |
| INST | Institutional (University) |
| GOV | Government bodies |